\documentclass[useAMS,usenatbib,a4paper,referee]{mn2e}
\usepackage{graphicx}
\usepackage{amsmath}
\usepackage{txfonts}
\usepackage{mathrsfs}

\newcommand{\be}{\begin{equation}}
\newcommand{\ee}{\end{equation}}
\newcommand{\bea}{\begin{eqnarray}}
\newcommand{\eea}{\end{eqnarray}}

\title[Photon Geodesics in FRW Cosmologies]{Photon Geodesics in FRW Cosmologies}
\author[Ojeh Bikwa, Fulvio Melia and Andrew Shevchuk]{Ojeh Bikwa,$^{1}$ Fulvio Melia$^{2}$\thanks{John Woodruff Simpson Fellow. 
E-mail: fmelia@email.arizona.edu} and Andrew Shevchuk$^{3}$\thanks{NSF Graduate Fellow.}\\
$^{1}$Amherst College, Amherst MA 01002, USA\\
$^{2}$Department of Physics, The Applied Math Program, and Department of Astronomy, 
The University of Arizona, Tucson AZ 85721, USA\\
$^{3}$Department of Astronomy, The University of Arizona, Tucson AZ 85721, USA}

\begin{document}

\date{}

\pagerange{\pageref{firstpage}--\pageref{lastpage}} \pubyear{2010}

\maketitle

\label{firstpage}

\begin{abstract}
The Hubble radius is a particular manifestation of the Universe's 
gravitational horizon, $R_{\rm h}(t_0) \equiv c/H_0$, the distance 
beyond which physical processes remain unobservable to us at the 
present epoch. Based on recent observations of the cosmic microwave 
background (CMB) with WMAP, and ground-based and HST searches for 
Type Ia supernovae, we now know that $R_{\rm h}(t_0)\sim$$13.5$ Glyr. 
This coincides with the maximum distance ($ct_0\approx 13.7$ Glyr) 
light {\it could} have traveled since the big bang. However, the 
physical meaning of $R_{\rm h}$ is still not universally understood 
or accepted, though the minimalist view holds that it is merely the 
proper distance at which the rate of cosmic recession reaches the speed 
of light $c$. Even so, it is sometimes argued that we can see light 
from sources beyond $R_{\rm h}$, the claim being that $R_{\rm h}$ lies 
at a redshift of only $\sim$$2$, whereas the CMB was produced at a much 
greater redshift ($\sim$$1100$). In this paper, we build on recent 
developments with the gravitational radius by actually calculating null 
geodesics for a broad range of FRW cosmologies, to show---at least in 
the specific cases we consider here, including $\Lambda$CDM---that no 
photon trajectories reaching us today could have ever crossed 
$R_{\rm h}(t_0)$. We therefore confirm that the current Hubble 
radius, contrary to a commonly held misconception, is indeed the 
limit to our observability. We find that the size of the visible 
universe in $\Lambda$CDM, measured as a proper distance, is 
approximately $0.45ct_0$.
\end{abstract}

\begin{keywords}
{cosmic microwave background, cosmological parameters, cosmology: observations,
cosmology: redshift, cosmology: theory, distance scale}
\end{keywords}

\section{Introduction}
The standard model of cosmology is confronted with several unpalatable coincidences,
suggesting that we do not yet have a fully consistent picture of the Universe's
dynamical expansion (see, e.g., Melia \& Shevchuk 2011). Part of the problem is that 
cosmological observations can only be interpreted from within the context of a 
pre-assumed model, and the data can be quite compliant. 

The $\Lambda$CDM (Cold Dark Matter with a cosmological constant $\Lambda$) model 
has been without peer in cosmology (see, e.g., Spergel et al. 2003, and Tegmark 
et al. 2004). For example, this model has been used with complementary measurements 
of the cosmic microwave background (CMB) radiation to determine that the Universe is 
flat, so its energy density $\rho$ is at (or very near) the ``critical" density 
$\rho_{\rm c}\equiv 3c^2H^2/ 8\pi G$. But among the many peculiarities of this 
description of the universe is the inference, based on current observations, that the 
density $\rho_{\rm d}$ of dark energy must itself be of order $\rho_{\rm c}$. (In the 
context of $\Lambda$CDM, the best fit to the WMAP data indicates that dark energy 
represents approximately $73\%$ of the total $\rho\approx \rho_{\rm c}$; see Spergel et al. 2003.) 
Dark energy is often thought to be the manifestation of the aforementioned cosmological 
constant, $\Lambda$, though no reasonable explanation has yet been offered as to why 
such a fixed, universal density ought to exist at this scale. It is well known that 
if $\Lambda$ is associated with the energy of the vacuum in quantum theory, it should 
have a scale representative of phase transitions in the early Universe---120 orders 
of magnitude greater than $\rho_{\rm c}$.

Many workers have attempted to circumvent these difficulties by proposing alternative forms of 
dark energy, including Quintessence (Ratra \& Peebles 1988; Wetterich 1988), which represents 
an evolving canonical scalar field with an inflation-inducing potential, a Chameleon field
(see, e.g., Mota \& Barrow 2004; Khoury \& Weltman 2004; Brax et al. 2004) in which the
scalar field couples to the baryon energy density and varies from solar system to cosmological 
scales, and modified gravity, arising out of both string motivated, or General Relativity 
modified actions (Capozziello et al. 2003; Nojiri \& Odintsov 2003; Carroll et al. 2004), 
which introduce large length scale corrections modifying the late time evolution of the 
Universe. The actual number of suggested remedies is far greater than this small, 
illustrative sample.

An equally perplexing puzzle with $\Lambda$CDM has been dubbed the ``coincidence problem," 
arising from the peculiar near-simultaneous convergence of the matter energy density 
$\rho_{\rm m}$ and $\rho_{\rm d}$ towards $\rho_{\rm c}$ in the present epoch. Though 
$\rho_{\rm m}$ and $\rho_{\rm d}$ are expected to change at different rates as the 
Universe expands (particularly if dark energy is a cosmological constant) they are 
nearly equal in the present epoch, implying that we live at a special time in cosmic 
history.

In a recent paper (Melia \& Shevchuk 2011), we proposed an explanation for yet 
another disturbing coincidence, having to do with the apparent equality of our gravitational
(or Hubble) radius $R_{\rm h}$ with the distance $ct_0$ light could have traveled since the big 
bang (in terms of the presumed current age $t_0$ of the Universe). This equality has received
some scrutiny in recent years (Melia 2003, 2007, 2009, Melia \& Abdelqader 2009, van Oirschot 
et al. 2010; see also Lima 2007 for a related, though unpublished, work). 

Unfortunately, there is still some confusion regarding the properties of $R_{\rm h}$
due to a misunderstanding of the role it plays in our observations. For example, it is 
sometimes suggested  (see, e.g., Davis \& Lineweaver 2004; van Oirschot et al. 2010) 
that sources beyond $R_{\rm h}(t_0)$ are observable today, which is
certainly not the case. We will therefore begin by elaborating upon what the 
gravitational radius $R_{\rm h}$ is---and what it is not. Though first defined 
in Melia (2007), an unrecognized form of $R_{\rm h}$ actually appeared in de 
Sitter's (1917) own account of his spacetime metric. And we will advance the 
discussion further by actually calculating photon trajectories for various 
well-studied Friedmann-Robertson-Walker (FRW) Cosmologies, demonstrating that 
the null geodesics reaching us at $t_0$ have never crossed $R_{\rm h}(t_0)$. 
Some come close, and in one case---the de Sitter model---they approach 
$R_{\rm h}$ asymptotically as $t$ recedes to our infinite past. Our conclusion 
in this paper will be that $R_{\rm h}(t_0)$ is a real limit to our observability 
at the present time $t_0$. 

\section{\bf The Gravitational (Hubble) Radius}
Standard cosmology is based on the Robertson-Walker metric for a spatially
homogeneous and isotropic three-dimensional space. In terms of the proper
time $t$ measured by a comoving observer, and the corresponding radial ($r$) and angular 
coordinates ($\theta$ and $\phi$) in the comoving frame, an interval $ds$ in this
metric is written as
\begin{equation}
ds^2=c^2\,dt^2-a^2(t)[dr^2(1-kr^2)^{-1}+r^2(d\theta^2+\sin^2\theta\,d\phi^2)]\;,
\end{equation}
where $a(t)$ is the expansion factor and the constant $k$ is $+1$ for a closed
universe, $0$ for a flat, open universe, or $-1$ for an open universe. 

In recent work (Melia 2007, 2009; Melia \& Abdelqader 2009), we demonstrated
the usefulness of examining properties of the metric in terms of both
co-moving coordinates $(ct,r,\theta,\phi)$ and observer-dependent 
coordinates $(cT,R,\theta,\phi)$, where $R$ is the so-called proper radius
$a(t)r$. Whereas $(ct,r,\theta,\phi)$ describe events in a frame ``falling'' freely 
with the cosmic expansion, the second set of coordinates are referenced to a 
particular individual who describes the spacetime relative to the origin at his 
location.\footnote{It is worth mentioning that prior to the introduction of
co-moving coordinates in the 1920's, the cosmological spacetime was actually
conventionally expressed in terms of these observer-dependent coordinates (see, 
e.g., de Sitter 1917.)}

For a flat universe ($k=0$), it is straightforward to show that with this definition of $R$, Equation~(1)
becomes
\begin{equation}
ds^2= \Phi\left[c\,dt + \left(\frac{R}{R_{\rm h}} \right)\Phi^{-1} 
dR  \right]^2 - \Phi^{-1}{dR^2}-R^2\,d\Omega^2\;,
\end{equation}
where the function
\begin{equation}
\Phi\equiv 1-\left(\frac{R}{R_{\rm h}} \right)^2
\end{equation}
signals the dependence of the metric on the proximity of the observation radius
$R$ to the gravitational radius $R_{\rm h}$. As shown in Melia \& Abdelqader 
(2009), the exact form of $R_{\rm h}$ depends on the constituents of the 
Universe. For example, in de Sitter space (which contains only a cosmological 
constant $\Lambda$), $R_{\rm h}=c/H_0$, in terms of the (time-invariant) 
Hubble constant $H_0$. It is equal to $2ct$ in a radiation-dominated universe 
and $3ct/2$ when the Universe contains only (visible and dark) matter.

Quite generally, $R_{\rm h}$ is the radius at which a proper sphere encloses sufficient 
mass-energy to turn it into a Schwarzschild surface for an observer at the origin 
of the coordinates (see Melia \& Shevchuk 2011). That is, $R_h$ is given in 
all cases by the expression
\begin{equation}
R_{\rm h}={2GM(R_{\rm h})\over c^2}\;,
\end{equation}
where $M(R_{\rm h})=(4\pi/3) R_{\rm h}^3\,\rho/c^2$, in terms of the total energy 
density $\rho$. Thus, $R_{\rm h}=({3c^4/8\pi G\rho})^{1/2}$ which, for a flat
universe, may also be written more simply as $R_{\rm h}(t)=c/H_0(t)$.

Although not defined in this fashion, the Hubble radius $c/H_0(t)$
(more commonly encountered when co-moving coordinates are used)
is therefore seen to coincide with the gravitational radius $R_{\rm h}$ 
emerging directly from the Robertson-Walker metric written in terms of 
$R$. What this means, of course, is that the Hubble radius is not a mere 
empirical artifact of the expanding universe, but actually represents
the radius at which a sphere centered on the observer contains sufficient
mass-energy for its surface to function as a {\it static} horizon.
Of course, that also means that the speed of expansion a distance $R_{\rm h}$
away from us must be equal to $c$, just as the speed of matter falling
into a compact object reaches $c$ at the black hole's horizon (and this is
in fact the criterion used to define the Hubble radius in the first place).

The reason why $R_{\rm h}$ is an essential ingredient of the metric 
written in the form of Equation~(2) can be understood in the context of
the Birkhoff (1923) theorem and its corollary (see Melia 2007). The 
latter states that the metric inside an empty spherical cavity, at the 
center of a spherically symmetric system, must be equivalent to the 
flat-space Minkowski metric. Space must be flat in a spherical cavity 
even if the system is infinite. Thus, the metric between the edge of
a cavity and a spherically symmetric mass placed at its center is 
necessarily of the Schwarzschild type. The worldlines linked to an 
observer in this region are curved relative to the center of the 
cavity in a manner determined solely by the mass we have placed 
there. The implication is that any two points within a medium with 
non-zero $\rho$ experience a net acceleration (or deceleration) toward
(or away) from each other, determined solely by how much mass-energy
is contained between them. This is why, of course, the Universe cannot 
be static, a concept that eluded Einstein himself, since his thinking on 
this subject preceded that of Birkhoff.

Yet the physical meaning of $R_{\rm h}$ is still elusive to many,
possibly because of the widely held belief that all horizons must necessarily 
be asymptotic surfaces attained when $t\rightarrow\infty$.
The so-called event horizon (see Rindler 1956) is indeed of this type,
representing the ultimate limit of our observability to the end of time.
However, $R_{\rm h}$ is not in this category---nor should it be. Unlike
the event horizon, the gravitational radius is a time-dependent 
quantity that increases in value at a rate determined by the evolving 
constituents of the Universe, specifically, the value of the equation-of-state 
parameter $w$, defined by the relation $p=w\rho$, where $p$ is the pressure 
and $\rho$ is the energy density. For some cosmologies, $R_{\rm h}$ 
may turn into the event horizon when the cosmic time approaches infinity.

Note from the definition of $R_{\rm h}$ that 
$\dot{R}_{\rm h}/R_{\rm h}=-\dot{\rho}/2\rho$, so with Equation~(4)
\begin{equation}
\dot{R}_{\rm h}={3\over 2}(1+w)c\;.
\end{equation}
Clearly, $R_{\rm h}$ is constant only for the de Sitter metric, where 
$w=-1$ and therefore $\dot{R}_{\rm h}=0$. For all other values of $w>-1$,
$\dot{R}_{\rm h}>0$. So in $\Lambda$CDM, for example, where the Universe
is currently dominated by a blend of matter and dark energy, $\dot{R}_{\rm h}>0$.
If dark energy were a cosmological constant, however, the Universe would eventually
become de Sitter as the density of matter drops to zero, and we would therefore
expect $R_{\rm h}$ to then asymptotically approach a constant value equal to the
radius of the event horizon in $\Lambda$CDM.

What this means, then, is that the current location of $R_{\rm h}$ affects 
what we can observe right now, at time $t_0$ since the big bang; it is not---and
is not meant to be---an indication of how far we will see in our future. One 
must therefore be careful interpreting spacetime diagrams, such as Figure~3 
in van Oirschot et al. (2010). Consider that for an Einstein-de Sitter universe 
(with $w=0$, the value used to construct this figure), $\dot{R}_{\rm h}=(3/2)c$ 
so our observations never encounter $R_{\rm h}$ because no matter how far light 
travels, it never overtakes $R_{\rm h}$. But that doesn't mean that this must 
always be the case. In fact, $\dot{R}_{\rm h}<c$ for any cosmology with 
$w<-1/3$---which is presumably the situation with $\Lambda$CDM. One of 
the principal goals of this paper is to calculate the actual photon
trajectories for well-known FRW cosmologies, such as $\Lambda$CDM, in
order to demonstrate how and why $R_{\rm h}$ functions as a limit to 
our observability today (at time $t_0$). 

\section{Null Geodesics in FRW Cosmologies}
Let us now derive the equation describing photon trajectories in
a cosmology consistent with the FRW metric (Equation~1). From the
definition of proper radius, we see that
\begin{equation}
\dot{R}=\dot{a}r+a\dot{r}\;.
\end{equation}
But the null condition for geodesics (see, e.g., Ellis \& Rothman 1993)
applied to Equation~(1) yields
\begin{equation}
c\,dt=-a(t){dr\over\sqrt{1-kr^2}}\;,
\end{equation}
where we have assumed propagation of the photon towards the
origin along a radius. The best indications we have today are
that the universe is flat so, for simplicity, we will assume
$k=0$ in all the calculations described below, and therefore
(for a photon approaching us)
$\dot{r}=-c/a$. Thus, we can write Equation~(6) as follows:
\begin{equation}
\dot{R}_\gamma=c\left({R_\gamma\over R_{\rm h}}-1\right)\;,
\end{equation}
where we have added a subscript $\gamma$ to emphasize the
fact that this represents the proper radius of a photon
propagating towards us. Note that in this expression, both
$R_\gamma$ and $R_{\rm h}$ are functions of cosmic time $t$.
The gravitational radius must therefore be calculated according
to Equation~(5).

The analysis of the WMAP data, within the context of the standard model, points
to an age for the Universe of $t_0\approx 13.7$ billion years. But we also know 
$\rho$ quite accurately now, and we can use it to calculate $R_{\rm h}(t_0)$, which 
appears to be approximately $13.5$ billion light-years, equal to $ct_0$ within 
the measurement errors. This is the unlikely coincidence we alluded to in
the introduction, because there clearly is no particular reason why 
$\dot{R}_{\rm h}$ in Equation~(5) should be equal to $c$, especially if 
$w$ changes with time.

The implausibility of this equality and its possible interpretation have been 
discussed elsewhere (e.g., Melia \& Shevchuk 2011), so we will not dwell on them here. However,
we will use the inferred value of $R_{\rm h}$ today as one of our boundary
conditions. In principle, Equation~(8) may be integrated either forwards
or backwards, but in reality, we are more familiar with the physical
conditions now, at time $t_0$, so it makes practical sense to
begin the solution of this equation at $t=t_0$, where $R_\gamma=0$
and $R_{\rm h}(t_0)=ct_0$. 

Not surprisingly, $\dot{R}_\gamma$ equals $-c$ today, since this must represent 
the actual speed of light measured in our local frame at the origin of 
our coordinates. Notice, however, that $\dot{R}_\gamma$ differs from
$-c$ away from the observer. This type of phenomenon often gives rise
to misinterpretation and confusion, stemming from the fact that many
consider $R$ to be the actual physical distance measured by a single 
observer, and that $dR/dt$ is therefore the physical speed. But in 
general relativity, the proper velocity measured by an individual is 
actually $v\equiv \sqrt{g_{RR}/g_{TT}}dR/dT$, calculated in terms of the 
proper distance $R$ and proper time $T$ in his/her frame, where the 
metric coefficients $g_{RR}$ and $g_{TT}$ are independent of
$T$. Light satisfies the null condition $ds=0$, and therefore
$v$ is {\it always} equal to $c$, regardless of whether the frame 
of reference is inertial or not. Written in terms of the co-moving 
coordinates, the metric in Equation~(1) does not satisfy these 
conditions, and so $\dot{R}_{\gamma}$ is generally not equal
to $c$. 

This happens because $t$ only represents the physical 
time on the clock at the observer's location (in other words,
$t=T$ only at $R=0$). If this observer were
to measure $T$ at any other position, (s)he would infer a value different 
from the local proper time $t$ at that radius. And because $t$ is used as 
a common time everywhere, the quantity $R(t)$ at best represents the sum 
of all the incremental segments measured by a collaboration of 
observers---each at time $t$ in their own rest frame---stretched out 
from one spacetime point to another (see, e.g., Weinberg 1972). Thus, 
although $R(t)$ is the ``proper" distance defined in terms of the
cosmic time $t$, it actually does not represent the physical distance 
between these two points for an individual observer whose time
coordinate is $T$ in his/her frame. A better description for $R$ 
would be that it represents a ``community" distance between two 
spacetime points and, as a result, $\dot{R}_\gamma$ is not 
constrained to have the value $-c$ away from the origin.

In the next section, we will discuss the solutions to Equation~(8) for
various assumed cosmologies, including $\Lambda$CDM, the current 
standard model.

\section{Results and Discussion}
The de Sitter universe has no ordinary matter or radiation. Its dynamics is 
dictated solely by a cosmological constant $\Lambda$, which results in an
expansion factor $a(t)=e^{H_0t}$. As we noted earlier (following
Equation~3), the gravitational horizon for this spacetime is constant
$R_{\rm h}=c/H_0$ and $w=-1$. By design, the energy density
$\rho$ is also constant. A plot of $R_\gamma$,
in units of $R_{\rm h}(t_0)$, is shown as a function of $t/t_0$ in
figure~1. In de Sitter, the maximum proper distance of photons
reaching us today was $\approx 0.65R_{\rm h}(t_0)$, which 
occurred at cosmic time $t=0$. Never along its geodesic path 
did light's proper distance from us exceed our gravitational 
horizon today. 

\begin{figure}
\begin{center}
{\includegraphics[scale=0.85,angle=0]{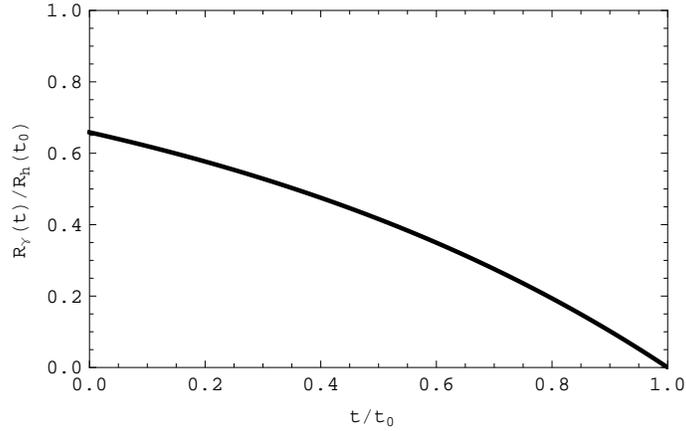}
\caption{The null geodesic, $R_\gamma$, as a function of cosmic
time, $t$, for de Sitter space. The proper radius is expressed in
units of the current gravitational radius $R_{\rm h}(t_0)$, and
$t$ is measured as a fraction of the Hubble time $t_0$.}}
\end{center}
\end{figure}

\begin{figure}
\begin{center}
{\includegraphics[scale=0.85,angle=0]{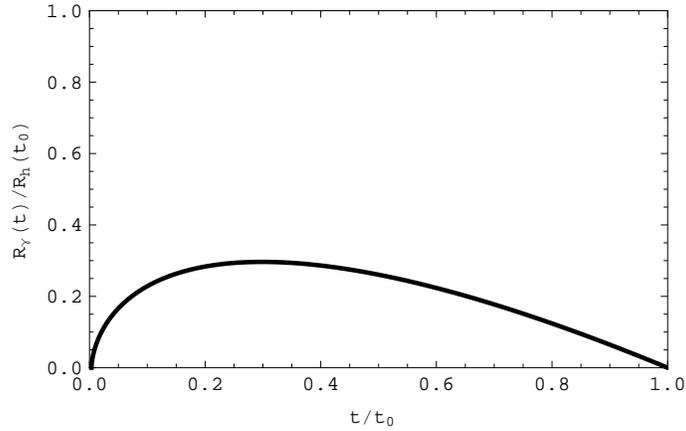}
\caption{Same as figure~1, except here for the Einstein-de Sitter
universe, consisting entirely of matter.}}
\end{center}
\end{figure}

This feature is echoed by the Einstein-de Sitter universe, which
consists entirely of matter. The cosmological constant and radiation 
are both absent, so $w=0$.  In this case, the scale factor grows as 
$a(t)=(3H_0t/2)^{2/3}$ (see, e.g., Melia \& Abdelqader 2009), 
so the density is infinite at $t=0$ and decreases monotonically
as $t\rightarrow t_0$. The gravitational radius ($R_{\rm h}=3ct/2$) grows linearly with 
time, but is always greater than $ct$. The null geodesic for this cosmology is shown in 
figure~2, where we see that $R_\gamma$ attains its maximum value $\sim$$0.3 R_{\rm h}(t_0)$ 
about 2.6 Gyr after the big bang. 

When the universal expansion is driven solely by radiation, $w=+1/3$ and
$R_{\rm h}=2ct$. The expansion factor is then given by $a(t)=(2H_0t)^{1/2}$, and 
the corresponding null geodesic is shown in figure~3. The proper distance 
$R_\gamma$ of light attains its maximum value of $\sim$$0.4 R_{\rm h}(t_0)$ at 
$t\approx 0.4t_0$.

\begin{figure}
\begin{center}
{\includegraphics[scale=0.85,angle=0]{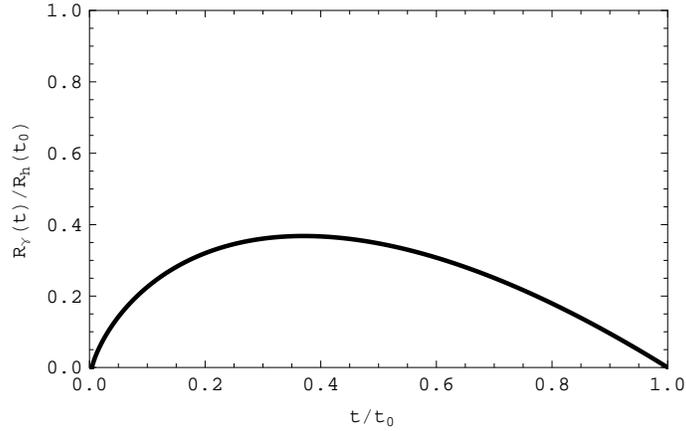}
\caption{Same as figure~1, except for a universe dominated by
radiation, for which $w=+1/3$ and $R_{\rm h}=2ct$.}}
\end{center}
\end{figure}

For a universe dominated by a combination of matter and radiation, the 
expansion factor also grows monotonically with cosmic time $t$, though 
the value of $w$ falls sharply 
from about $+1/3$ in the early universe (where radiation dominates) and 
tapers off towards zero as we approach the present time, where the effect
of radiation is relatively insignificant. The photon geodesic for this case
is shown in figure~4, where we see it attaining its maximum proper distance
from us, 
$\sim$$0.3 R_{\rm h}(t_0)$, roughly 3 Gyr after the big bang. As with
the other cases we've considered, the photon's trajectory never takes it
beyond our current gravitational horizon $R_{\rm h}(t_0)$.

\begin{figure}
\begin{center}
{\includegraphics[scale=0.85,angle=0]{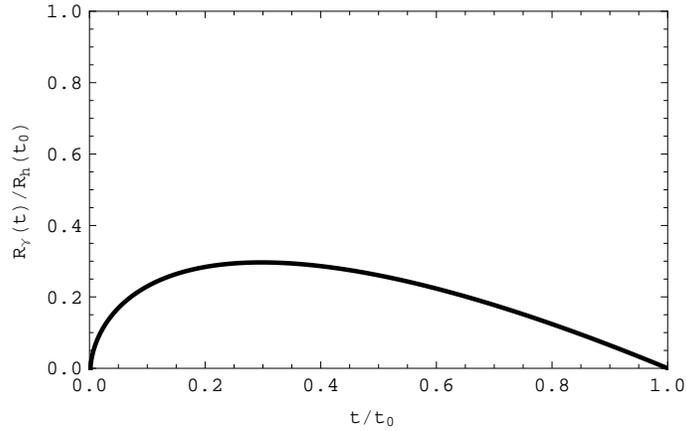}
\caption{Same as figure~1, except for a universe dominated by
matter and radiation.}}
\end{center}
\end{figure}

\begin{figure}
\begin{center}
{\includegraphics[scale=0.85,angle=0]{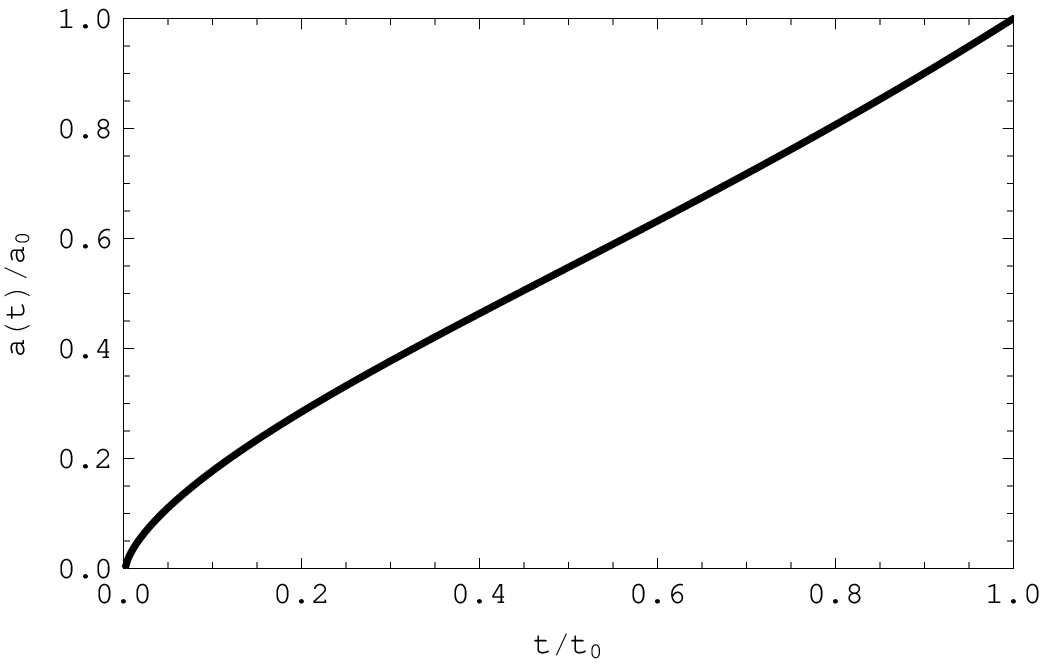}
\center{{\bf Figure 5a.} The expansion factor $a(t)$ as a 
function of cosmic time $t$ for the standard model of
cosmology, $\Lambda$CDM. The constant $a_0$ represents
the value of $a$ at the present time (which in a flat
cosmology may also simply be set to $1$).}}
\end{center}
\end{figure}

In $\Lambda$CDM, the standard model of cosmology, the density $\rho$
is comprised of three principal components,
\begin{equation}
\rho=\rho_{\rm m}+\rho_{\rm r}+\rho_{\rm de}
\end{equation}
where, following convention, we designate the matter, radiation, and dark
energy densities, respectively, as $\rho_{\rm m}$, $\rho_{\rm r}$, and
$\rho_{\rm de}$. We will also assume that these energy densities scale
according to $\rho_{\rm m}\propto a^{-3}$, $\rho_{\rm r}\propto a^{-4}$,
and $\rho_{\rm de}\propto f(a)$. (If dark energy is indeed a cosmological
constant $\Lambda$, which we assume for this calculation, then $f(a)=$ constant.) 
In this case, there is no simple analytical solution for $w$ and $R_{\rm h}(t)$,
so these are calculated numerically, along with $R_\gamma(t)$. For completeness,
we show the expansion factor $a(t)$ as a function of cosmic time in figure~5a;
the value of $w$, averaged over time from $0$ to $t$, is shown as a function
of cosmic time in figure~5b; the gravitational radius $R_{\rm h}$ is shown
as a function of $t$ in figure~5c; and the null geodesic itself is plotted
in figure~5d. The blending of three different components in $\rho$ produces
some discernible differences in the behavior of $R_\gamma$ compared to
the cases we considered previously, but even here, its maximum value was 
a fraction $\sim$$0.45$ of today's gravitational radius $R_{\rm h}(t_0)$ at
roughly $0.3t_0$ after the big bang.

\begin{figure}
\begin{center}
{\includegraphics[scale=0.85,angle=0]{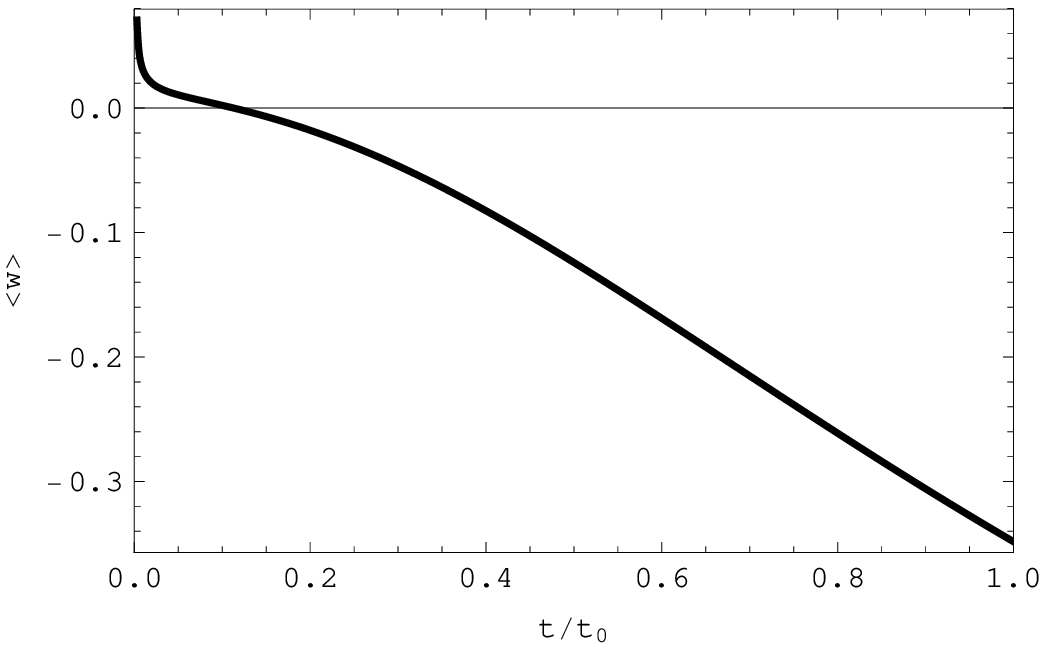}
\center{{\bf Figure 5b.} Same as figure~5a, except here for
the value of $w$ time-averaged from $0$ to $t$. Early on,
when radiation dominates the equation of state, $\langle
w\rangle$ is approximately $+1/3$, but it tapers off and
eventually reaches the value $-1/3$ at the present time.}}
\end{center}
\end{figure}

\begin{figure}
\begin{center}
{\includegraphics[scale=0.85,angle=0]{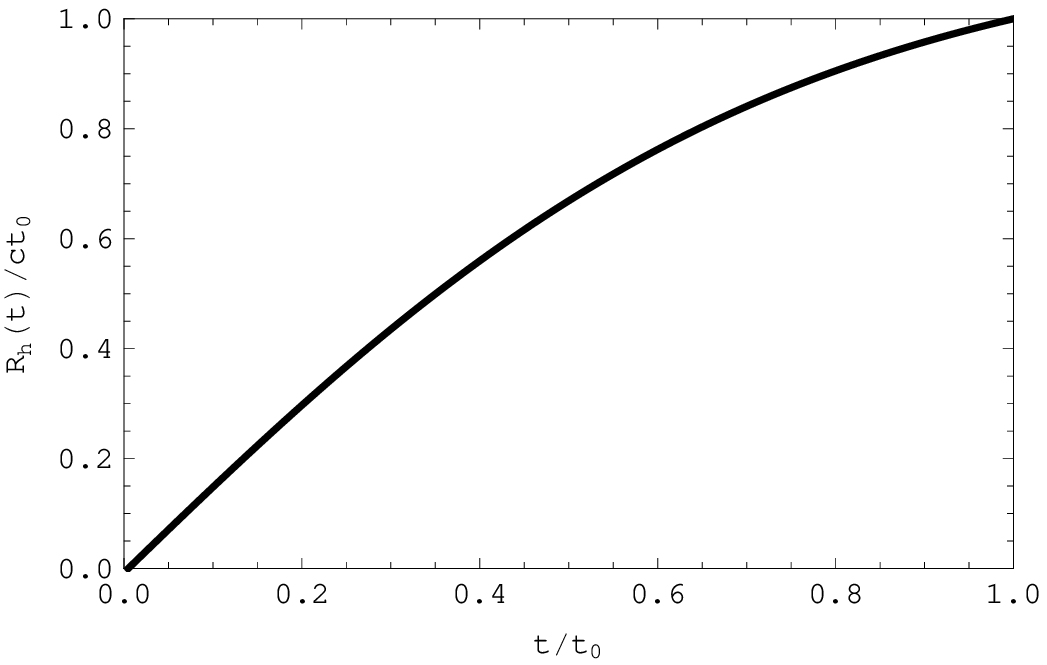}
\center{{\bf Figure 5c.} Same as figure~5a, except here
for the gravitational radius $R_{\rm h}$ as a function
of $t$.}}
\end{center}
\end{figure}

\begin{figure}
\begin{center}
{\includegraphics[scale=0.85,angle=0]{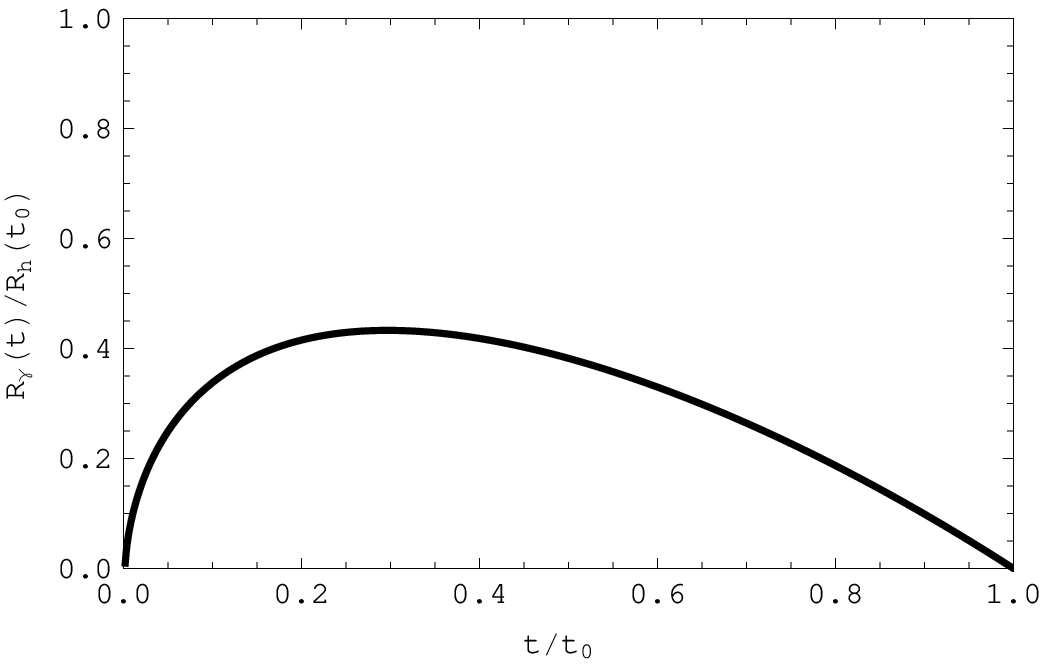}
\center{{\bf Figure 5d.} The null geodesic as a function
of cosmic time $t$ for $\Lambda$CDM.}}
\end{center}
\end{figure}

In all these cases, we have demonstrated through direct computation that null geodesics 
have never traversed the current gravitational (or Hubble) radius, regardless of what
specific cosmology one assumes. In order for light to reach us today, its source 
radiating at cosmic time $t_e$ must have been located a proper distance $R(t_e)$ 
away, corresponding to one of the trajectories shown in figures~1,2,3,4, or 5d. In 
other words, none of the sources whose light we detect today could have been 
more distant than (in most cases, actually only a fraction $\sim$0.3--0.4 of) our 
current gravitational horizon $R_{\rm h}(t_0)$.

In spite of this, it is sometimes claimed (see, e.g., van Oirschot, Kwan
and Lewis 2010) that we see sources from beyond the Hubble radius because
its redshift is only $\sim$$2$, much smaller than sources such as the recombination
region that produced the CMB (at a redshift of $\sim$$1100$ in the standard model).
This misconception arises because---written as a proper distance (see Melia \&
Shevchuk 2011)---$R_{\rm h}$ is expressed as
\begin{equation}
R_{\rm h}(t) = a(t)r_{\rm h}\;,
\end{equation}
where the comoving radius may further be written in the form
\begin{equation}
r_{\rm h}=c\int^t_{t_e}{dt'\over a(t')}\;.
\end{equation}
So, for example, if we crudely assume that the Universe's expansion has
been driven by a matter-dominated $\rho$, for which $a(t)=t^{2/3}$, we
find that
\begin{equation}
R_{\rm h}(t_0)\approx 3ct_0-3c{t_0}^{2/3}{t_e}^{1/3}\;,
\end{equation}
and using the observation that $R_{\rm h}(t_0)\approx ct_0$ (Melia
\& Shevchuk 2011), with 
\begin{equation}
1+z={a(t_0)\over a(t_e)}
\end{equation}
(see, e.g., Weinberg 1972), one easily ``finds" a redshift $z_{\rm h}$$\sim$$1.25$ for the Hubble
sphere (the value is closer to $2$ when using the actual form of $a(t)$ from 
$\Lambda$CDM).

The apparent conflict between this inference and the curves shown in
figures~1,2,3,4, and 5d, arises from an incorrect interpretation of $R_{\rm h}(t)$ 
as a null geodesic, which it is not. In other words, the comoving radius $r_{\rm
h}$ in Equation~(11), and therefore the proper radius $R_{\rm h}$, do not
track the propagation of light through the Hubble flow. Only Equation~(8) for
$R_\gamma(t)$ can do this because it includes both the effects of an expanding
medium and the change in comoving radius $r$ as light approaches us. Instead,
$R_{\rm h}(t_0)$ represents today's proper distance to sources that in the
past radiated the light we see at $t_0$ redshifted by an amount $z_{\rm h}$.
Therefore, to correctly determine their proper distance at the time the light
was produced, one needs to calculate $R_{\rm h}(t_e)$, not $R_{\rm h}(t_0)$, 
where $t_e$ is the cosmic time corresponding to redshift $z_{\rm h}$. Clearly,
$z_{\rm h}$ is that special redshift for which $R_{\rm h}(t_e)=R_\gamma(t_e)$,
and there is therefore no conflict between the finite value of $z_{\rm h}$ and
the fact that $R_{\rm h}(t_0)$ is the limit to our observability today.

\section{Concluding Remarks}
Throughout this paper, we have made a conscious effort to discuss the
properties of null geodesics in FRW cosmologies without resorting to 
conformal diagrams. This approach, also used to great effect by
Ellis and Rothman (1993), makes it easier to think in terms of familiar
quantities (proper distances and proper time) that are not always
straightforward to interpret otherwise. Misconceptions often arise
from the misinterpretation of coordinate-dependent effects. In their
paper, Ellis and Rothman clearly delineated true horizons from apparent
horizons, and extended the definitions, first introduced by Rindler (1956),
in a clear and pedagogical manner. In this paper, we have paid particular
attention to the gravitational horizon, also manifested as the Hubble
radius, which is time-dependent and may or may not turn into an event
horizon in the asymptotic future, depending on the equation of state.

It is useful at this point to be absolutely clear about how far light has traveled
in reaching us. It is quite evident from our results that sources whose light
we see today were at a proper distance $R(t_e)\rightarrow 0$ when 
$t_e\rightarrow 0$. For the FRW cosmologies we have considered here
(which include $\Lambda$CDM) the light reaching us today---including that
from the recombination region associated with the CMB---has traveled a {\it net}
proper distance of at most $\sim$0.3--$0.4ct_0$. It is therefore not correct
to claim that the size of the visible universe in these cosmologies is $ct_0$ 
(or even greater in some interpretations). Because all causally connected 
sources in an expanding universe began in a vanishingly small volume as 
$t\rightarrow 0$, the maximum proper distance from which we receive light 
today must necessarily be less than $ct_0$, since presumably there were 
no pre-existing sources at a non-zero proper distance prior to $t=0$ with 
which we were in causal contact.

It is true, however, that more and more sources become visible to us
as time advances, since for $t>t_0$, the geodesic curves terminating
in our future all rise above their current counterparts shown 
in figures~1,2,3,4, and 5d. In our future, we will see light from 
sources that radiated at proper distances greater than those
shown here. Of course, $R_{\rm h}(t)$ will also continue to 
increase, and it is not difficult to convince oneself from 
Equations~(5) and (8) that the limits of observability will always
be $R_{\rm h}(t)$, since $R_\gamma(t_e)/R_{\rm h}(t)$ is
smaller than $1$ for all $t_e<t$. To see this, let us first 
find the emission time $t_{e,\; {\rm max}}$ at which 
$R_\gamma(t_e)$ attains its maximum value.  Clearly, this happens
when $\dot{R_\gamma}=0$, which means that $R_\gamma(t_{e,\;{\rm max}})
=R_{\rm h}(t_{e,\;{\rm max}})$. But for all $w>-1$, Equation~(5) shows that
$\dot{R_{\rm h}}>0$, and therefore $R_{\rm h}(t)>R_{\rm h}(t_{e,\;{\rm max}})$
for $t>t_{e,\;{\rm max}}$, which also means that $R_{\rm h}(t)> 
R_\gamma(t_{e,\;{\rm max}})$ for all $t>t_{e,\;{\rm max}}$.

\section*{Acknowledgments}
This research was partially supported by ONR grant N00014-09-C-0032
and an NSF Graduate Fellowship at the University of Arizona, and by a Miegunyah Fellowship at the
University of Melbourne. We are particularly grateful to Amherst College
for its support through a John Woodruff Simpson Lectureship. Finally,
we wish to thank Roy Kerr for many helpful and
enjoyable discussions.

\label{lastpage}
\end{document}